\documentclass[fleqn,12pt,twoside]{article}
\usepackage{espcrc1}
\usepackage{epsfig,amssymb}
\long\def\@makefigurecaption#1#2{#1. #2\par}

\newcommand{\AmS}{{\protect\the\textfont2
  A\kern-.1667em\lower.5ex\hbox{M}\kern-.125emS}}
\title{Hadronic interaction of $\eta$ and $\eta^{\prime}$ mesons with 
       nucleons\thanks{This work has been partly supported by the European Community -- Access
       to Research Infrastructure action of the Improving Human Potential
       Programme.}}
 \author{P.~Moskal\address[fz]{IKP \& ZEL Forschungszentrum J\"{u}lich, D-52425 J\"{u}lich,
  Germany}\address[uj]{Institute of Physics, Jagellonian University, PL-30-059 Cracow, Poland},
 H.-H.~Adam\address[mue]{IKP, Westf\"{a}lische Wilhelms--Universit\"{a}t, D-48149 M\"{u}nster, Germany},
 A.~Budzanowski\address[ifj]{Institute of Nuclear Physics, PL-31-342 Cracow, Poland},
 R.~Czy{\.{z}}ykiewicz\addressmark[uj], 
 D.~Grzonka\addressmark[fz],
 M.~Janusz\addressmark[uj],
 L.~Jarczyk\addressmark[uj],
 B.~Kamys\addressmark[uj],
 A.~Khoukaz\addressmark[mue],
 K.~Kilian\addressmark[fz],
 P.~Kowina\addressmark[fz]\address[us]{Institute of Physics,
                       University of Silesia, PL-40-007 Katowice, Poland},
 T.~Lister\addressmark[mue],
 W.~Oelert\addressmark[fz],  
 T.~Ro{\.{z}}ek\addressmark[fz]\addressmark[us],
 R.~Santo\addressmark[mue],
 G.~Schepers\addressmark[fz],
 T.~Sefzick\addressmark[fz],
 M.~Siemaszko\addressmark[us],
 J.~Smyrski\addressmark[uj],
 S.~Steltenkamp\addressmark[mue],
 A.~Strza{\l}kowski\addressmark[uj],
 P.~Winter\addressmark[fz],
 M.~Wolke\addressmark[fz]\thanks{present address: The Svedberg
      Laboratory, Thunbergsv\"agen 5A, Box 533, S--75121 Uppsala, Sweden.},
 P.~W{\"u}stner\addressmark[fz],
 W.~Zipper\addressmark[us]
}
       
\begin{document}
\maketitle

\begin{abstract}
 Due to their short life-time,
 flavour-neutral mesons cannot be utilized
 as  free secondary beams or targets, and therefore a study of their 
 interaction with nucleons is not possible via  direct 
 scattering experiments.
 This interaction is, however,
   accessible via  its influence
 on the energy dependence --  and on the phase space distributions 
 of the cross sections for reactions in which these mesons 
 are produced.   

 In case of the $pp\to pp\eta$ reaction 
 the experimentally determined distributions of the differential cross sections
 close to the production threshold
 cannot be described by
 taking into account the S-wave proton-proton and proton-$\eta$ interaction only.  
 Here we show that the angular distributions determined at the COSY-11
 facility reveal some evidence 
 for P-wave admixture in the proton-proton subsystem 
 already at an excess energy as low as Q~=~15.5~MeV. We also present 
 that one can estimate the relative strength
 of the $\eta$-nucleon and $\eta^{\prime}$-nucleon interactions
 by comparison of the $\eta$ and $\eta^{\prime}$ production yield.
\end{abstract}

\section{COMPARISON OF THE p-$\eta$ 
         AND p-$\eta^{\prime}$ INTERACTIONS}
\label{}
Close to the kinematical threshold the total cross section
for the meson production via the nucleon-nucleon interaction
grows rapidly with increasing excess energy~Q.  
It is  well established~\cite{review} that for the 
$pp\to pp\eta$~\cite{eta_data}
and 
$pp\to pp\eta^{\prime}$~\cite{etap_data} reactions 
this total cross section
changes by about two orders of magnitude within a Q range of about ten MeV.
The shape of the excitation function is predominantly
determined by the 
changes of the phase space volume 
and by the final state interaction among the produced particles. 
The precision of the experiments performed at the cooler synchrotrons
allows to distinguish the subtle effects originating from the meson-nucleon 
interaction.
A quantitative derivation of the 
$p-\eta$  and $p-\eta^{\prime}$ hadronic potentials requires, however,
a sophisticated theoretical treatment 
since the distortion caused by the nucleons is by orders of
magnitude larger than that due to the meson--nucleon forces,
and  even small
fractional inaccuracies in the description of nucleon--nucleon effects may
obscure the inference on the meson--nucleon interaction.
To minimize the ambiguities which  may result from these
discrepancies 
-- at least
for the qualitative estimation of the effects of the unknown meson--nucleon
interaction --
 one can compare the spectra from the production of a meson under investigation 
to the spectra determined for the production of a meson whose interaction with
nucleons is established.
To visualize the influences 
of the $p-\eta$  and $p-\eta^{\prime}$ interaction on the energy dependence 
of the total cross section we have compared  
the modulus of the primary transition amplitude $|M_0|$
of the $pp\to pp\eta$ and $pp\to pp\eta^{\prime}$ reactions
to the one extracted from the data on the $pp\to pp\pi^0$ 
reaction~\cite{swave}~\footnote{
The S-wave $\pi$-proton interaction is negligibly weak
in comparison to the proton-proton one. The real part of the 
$\pi - p$ scattering length ( $|a_{p\pi}|~=~0.13$~fm~\cite{sigg269} ) 
is more than a factor of 50
smaller than $|a_{pp}|~=~7.83$~fm~\cite{naisse506}.}.
\begin{figure}[h]
 \parbox{0.5\textwidth}{
 \unitlength 1.0cm
  \begin{picture}(4.0,6.5)
    \put(0.0,0.0){
       \epsfig{figure=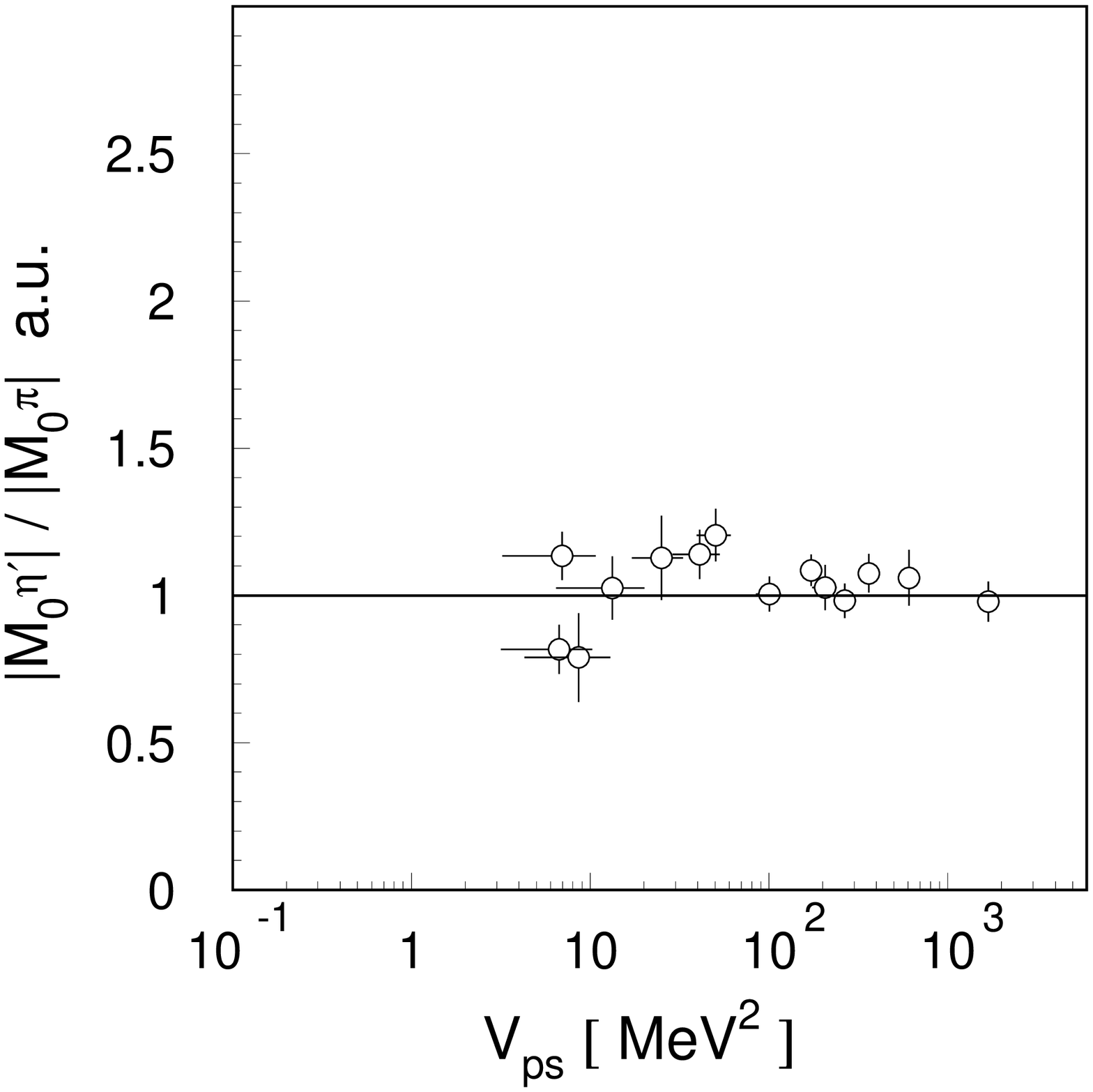,height=4.0cm,width=4.8cm,angle=0}
    }
       \put(5.0,1.5){
          { b)}
       }
    \put(0.0,3.1){
       \epsfig{figure=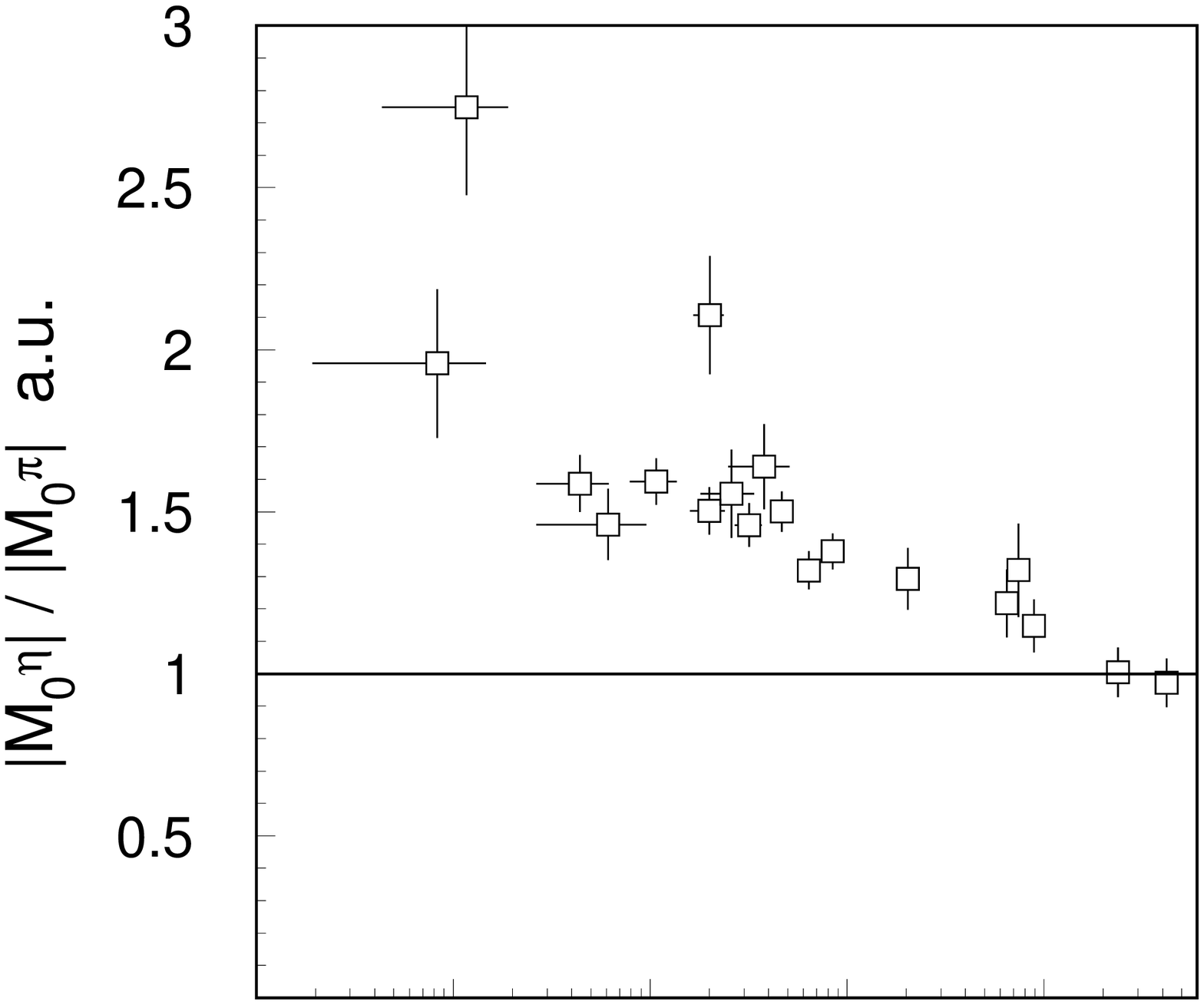,height=4.0cm,width=4.8cm,angle=0}
    }
       \put(5.0,6.0){
          { a)}
       }
  \end{picture}
 }
 \parbox{0.5\textwidth}{
  \caption{ 
            The ratios of \ \ a) $|M^{\eta}_{0}|/|M^{\pi^{0}}_{0}|$
             and \ \ b) $|M^{\eta^{\prime}}_{0}|/|M^{\pi^{0}}_{0}|$
             extracted from the experimental data for  $pp\to pp\eta$~\cite{eta_data}
             and $pp\to pp\eta^{\prime}$~\cite{etap_data} reactions.
             $|M^{\pi^{0}}_{0}|$
             was calculated by interpolating  the data of reference~\cite{pi0_data}.
             The figure is adapted from reference~\cite{swave}.
             \label{ratio}
        }
  }
\end{figure}

\vspace{-0.8cm}

Figures~\ref{ratio}a and~\ref{ratio}b show the
dependence of $|M_0|$ on the phase--space volume for $\eta$ and
$\eta^{\prime}$ production normalized to $|M_0^{\pi^0}|$.
The values of $|M_0|$ were extracted from the experimental data
disregarding the proton--meson interaction.
If the influence of the neglected interactions were the same in the case of
the $\eta~(\eta^{\prime})$  and $\pi^0$ production
the points would be consistent with the
solid line.
This holds in case of the $pp \rightarrow pp \eta^{\prime}$ reaction
indicating the weakness of the proton--$\eta^{\prime}$ interaction
independently of the prescription used for the proton--proton FSI~\cite{swave}.
In case of the $\eta^{\prime}$ meson its low--energy interaction with the
nucleons was expected to be very weak since there exists no baryonic resonance
which would decay into $N \eta^{\prime}$ channel~\cite{groom1}.
In contrary, the existence of the $N^*(1535)$ resonance, 
which decays significantly into  nucleon and the $\eta$ meson,
indicates that  
the $N \eta$ interaction is much stronger than the $N\eta^{\prime}$ one,
and indeed as depicted in figure~\ref{ratio}a 
the strong effects of the
$\eta pp$ FSI at low $V_{ps}$ are visible. 

As a next step for a quantitative understanding 
of the $pp\eta$ dynamics a full three-body description
of the system with the complex hadronic potentials is required
as well as 
an exact determination of the magnitudes 
of the contributing partial waves. Some aspects of the latter issue
are discussed in the next section.

\section{DIFFERENTIAL CROSS SECTIONS FOR THE REACTION $pp\to pp\eta$}
\label{section2}
\begin{figure}[h]
\parbox{0.35\textwidth}{\epsfig{file=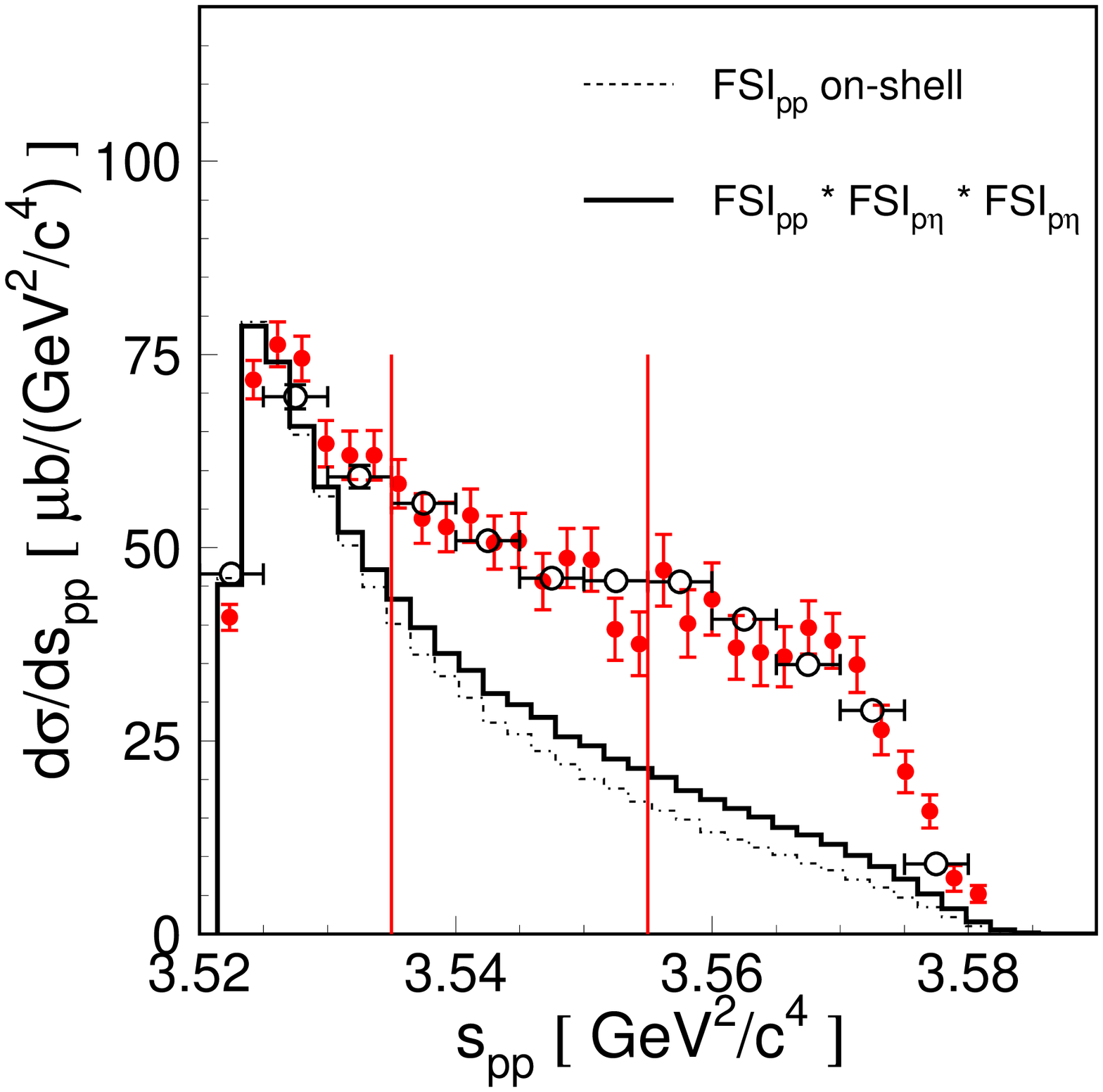,width=0.35\textwidth}} \hfill
\parbox{0.65\textwidth}{\epsfig{file=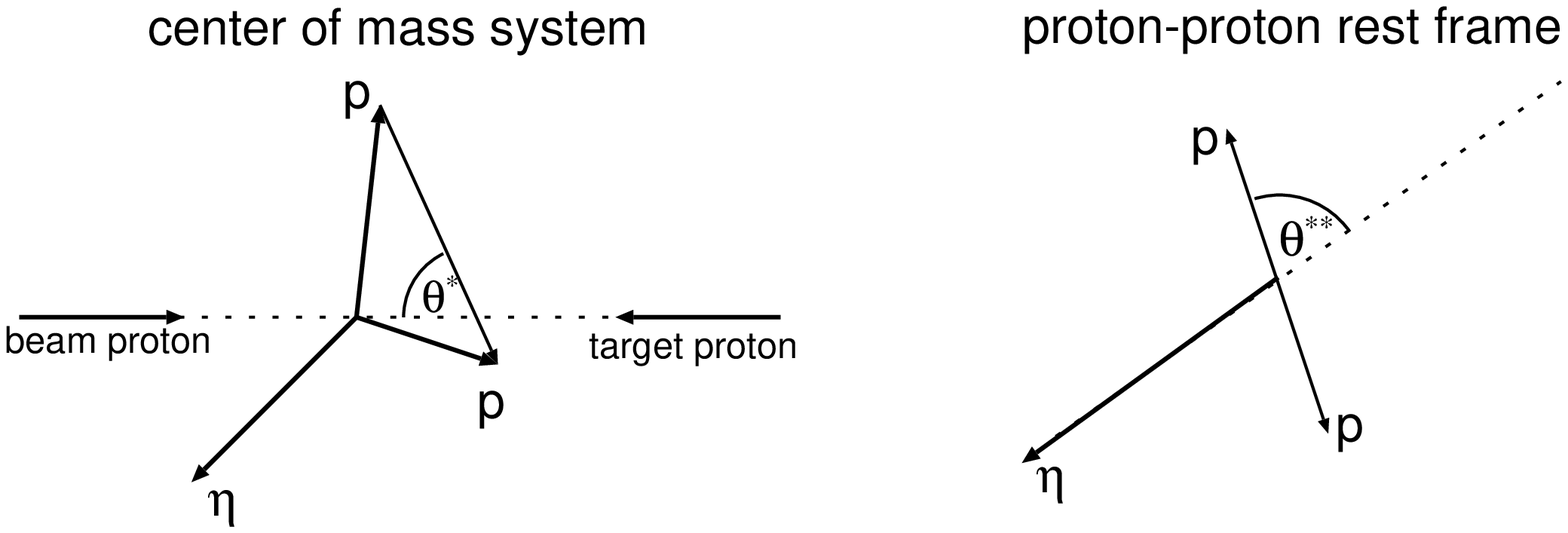,width=0.65\textwidth}} 

\vspace{-1cm}

   \caption{\label{figure2}
    {\bf (left side)} Distribution of the square of the proton-proton invariant mass ($s_{pp}$)
    for the $pp\to pp\eta$ reaction at an excess energy of Q~=~15.5 MeV.
    The data obtained by the TOF collaboration~\cite{TOFeta}(open circles)
    have been normalized in magnitude to the results of the COSY-11 
    collaboration~\cite{mesonqnp}(closed circles).
    The integrals of the phase space weighted by  
    the square of the proton-proton on-shell 
    scattering amplitude~(dashed line)--FSI$_{pp}$, and by the product of FSI$_{pp}$ and 
    the square of the proton-$\eta$ scattering amplitude~(solid line),    
    have been normalized arbitrarily at small values of $s_{pp}$.
    The solid line was obtained assuming a scattering length of
   $a_{p\eta}$~=~0.7~fm~+~$i$~0.4~fm. 
   {\bf (middle and right side)} Definition of the angles used in the text.
     }
\end{figure}

 On  previous conferences we have already reported on 
 the phase space density distribution determined 
 for the $pp\to pp\eta$ reaction at an
 excess energy of Q~=~15.5~MeV~\cite{menu,mesonqnp}.
 The obtained spectrum --~shown here in figure~\ref{figure2}~--
 revealed a strong deviation from the expectation based on the    
 factorization of the transition amplitude
 into the constant primary production
 and the on-shell incoherent pairwise interaction
 among the produced particles (see solid and dashed lines).

\vspace{-0.9cm}

\vspace{-0.8cm}

 The experimental data obtained independently by TOF~\cite{TOFeta} and COSY-11~\cite{mesonqnp}
 using different
 detection systems agree perfectly with each other and make  possible 
 systematical errors rather improbable. The data show also
 a fully isotropic distribution over the polar emission angle
 of the $\eta$ meson in the center-of-mass frame~\cite{TOFeta,menu},
 and are consistent with an isotropic angular distribution of
 the relative momentum 
 of the protons seen in the center-of-mass system~\cite{TOFeta}
 (angle $\theta^{*}$ on the middle side of figure~\ref{figure2}).

\begin{figure}[hbt]
\parbox{0.32\textwidth}{\epsfig{file=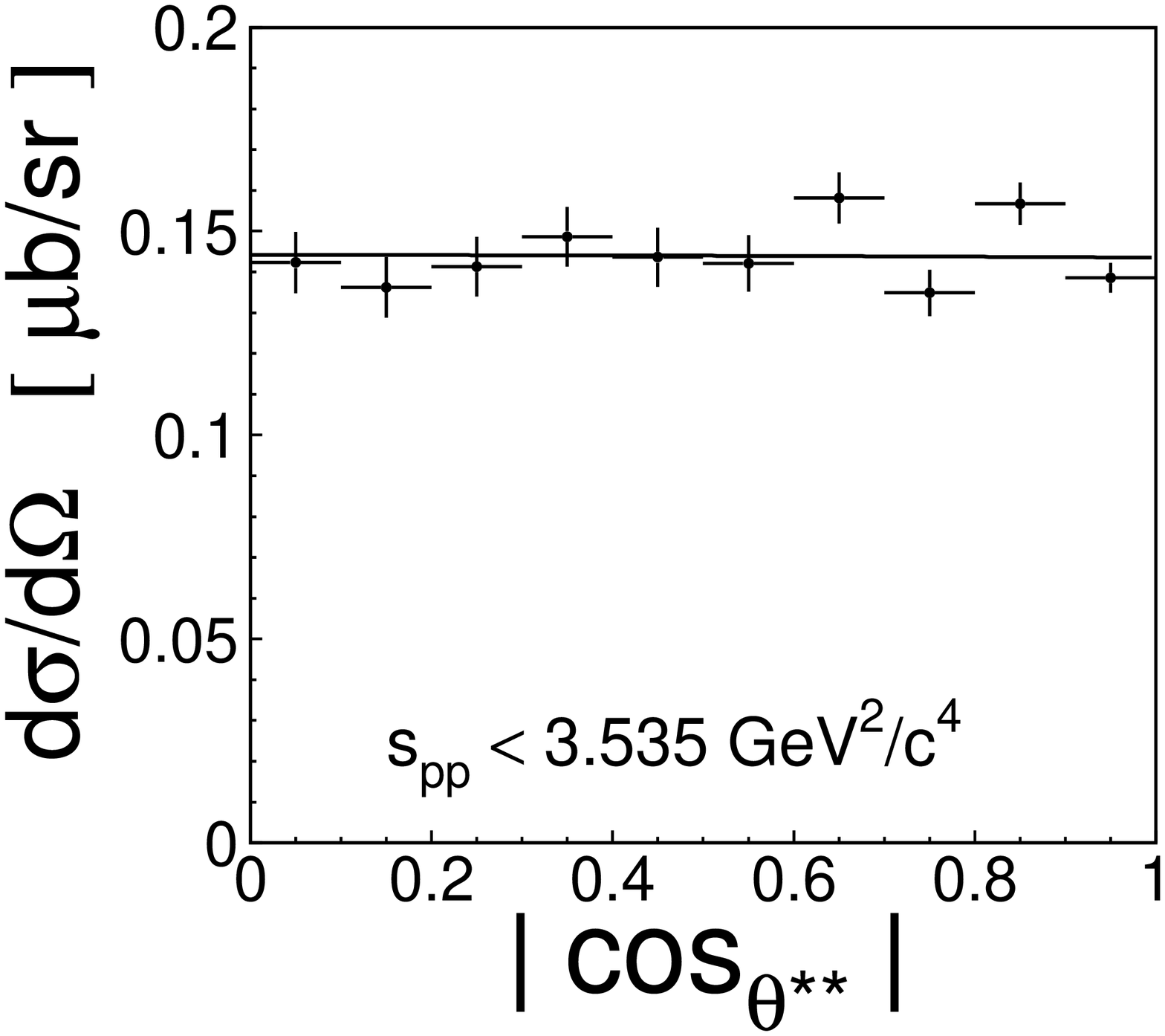,width=0.35\textwidth}} 
\parbox{0.32\textwidth}{\epsfig{file=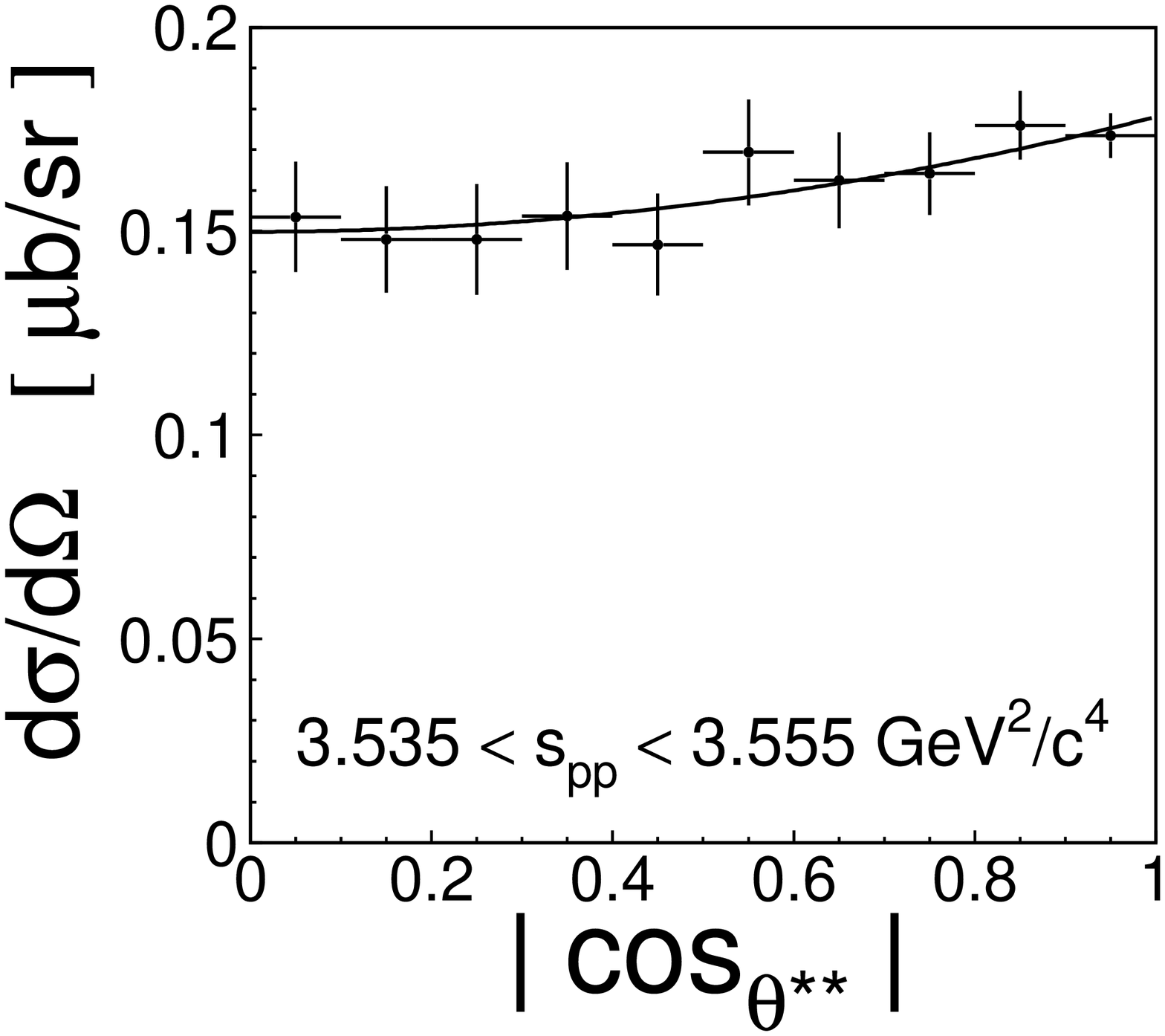,width=0.35\textwidth}}
\parbox{0.32\textwidth}{\epsfig{file=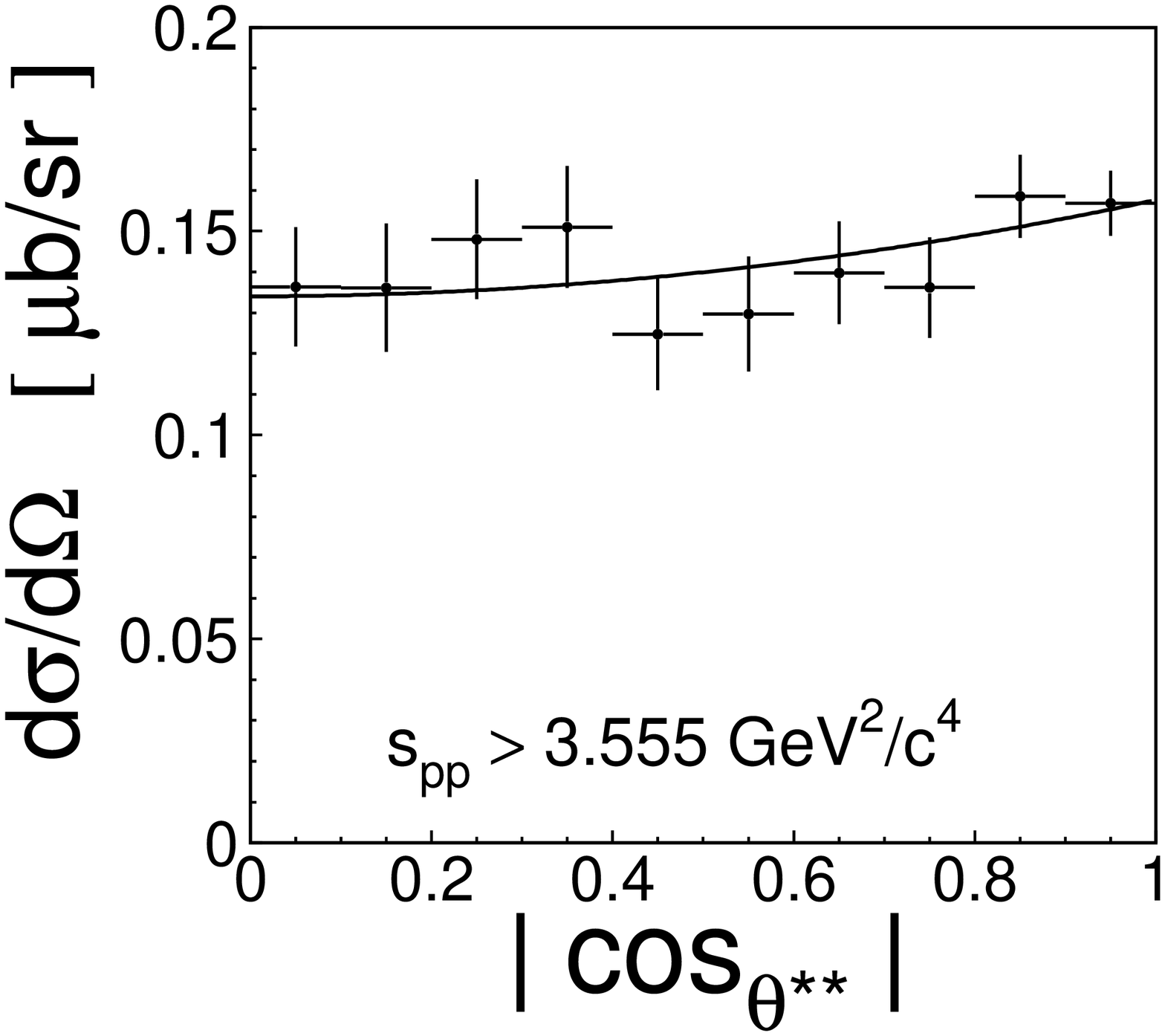,width=0.35\textwidth}}

\vspace{-1cm}

   \caption{\label{figure3} Distribution of the polar angle of the relative
    proton-proton momentum with respect to the momentum of the $\eta$ meson
    as seen in the di-proton rest frame. Figures correspond to the three different
    $s_{pp}$ intervals marked by the vertical lines in figure~\ref{figure2}.}
\end{figure}

 According to preliminary investigations based on the 
 meson exchange model~\cite{teoretycy}, 
 the observed distribution of the proton-proton invariant mass
 cannot be explained assuming that the production of the 
 outgoing particles takes place exclusively 
 with the relative angular momentum equal 
 to zero. The discrepancy between the solid
 line of figure~\ref{figure2} and the data is also too large to 
 be explained by the underestimation of the s-wave proton-$\eta$
 interaction. Inspired by that difficulty we checked the partial wave
 distribution in the proton-proton system deriving from the 
 data the  angular distribution of the proton momentum
 in the rest frame of proton-proton 
 system (angle $\theta^{**}$ on the right side of figure~\ref{figure2}).
  A possible non-zero angular momentum
  between outgoing protons should manifest itself in an unisotropic 
  population 
 of the angle between the relative proton-proton momentum
 and the recoil particle~($\eta$) seen from the 
 di-proton rest system~\cite{grzonkakilian}. 
 The distributions determined for three
 intervals of  $s_{pp}$ are shown in figure~\ref{figure3}.
 As a first step we restricted the analysis assuming that only 
 S-- and P-- waves contribute, and we fit the data by the linear
 combination of the Legendre polynomials up to the second degree,
which in case of two identical particles reads:
  $\displaystyle{ \frac{d\sigma}{d\Omega}~=~a~\left(~1~+~b~P_{2}(cos\theta^{**})~\right) }$.
In this representation the parameter $b$ is a measure of the relative amplitude 
of the P-- and S--wave contributions. 

\vspace{-1.1cm}

\vspace{-0.8cm}

As a result we obtained that $b$ is consistent
with zero for $s_{pp}~<~3.535$~GeV$^2$/c$^4$ and amounts 
to $\approx0.12$ for the middle and upper ranges of $s_{pp}$. This value 
supports the hypothesis that an admixture of P-waves in the proton-proton subsystem
is not negligible already at an excess energy of Q~=~15.5~MeV.

\end{document}